\documentclass[10pt,twocolumn]{article}
\usepackage[a4paper,margin=1.9cm]{geometry}
\usepackage{lmodern}
\usepackage[T1]{fontenc}
\usepackage[utf8]{inputenc}
\usepackage{microtype}
\usepackage{amsmath,amssymb,xcolor} 
\usepackage{booktabs}
\usepackage{hyperref}
\usepackage{natbib}
\usepackage{enumitem}
\usepackage{listings}
\usepackage{float}
\usepackage{caption}
\usepackage{algorithm}
\usepackage{algpseudocode}
\usepackage{tabularx}           
\newcolumntype{L}{>{\raggedright\arraybackslash}X}
\usepackage{graphicx}
\usepackage{animate}
\usepackage[most]{tcolorbox}
\usepackage[switch]{lineno}  

\newcommand*\patchAmsMathEnvironmentForLineno[1]{%
  \expandafter\let\csname old#1\expandafter\endcsname\csname #1\endcsname
  \expandafter\let\csname oldend#1\expandafter\endcsname\csname end#1\endcsname
  \renewenvironment{#1}%
    {\linenomath\csname old#1\endcsname}%
    {\csname oldend#1\endcsname\endlinenomath}}%
\newcommand*\patchBothAmsMathEnvironmentsForLineno[1]{%
  \patchAmsMathEnvironmentForLineno{#1}%
  \patchAmsMathEnvironmentForLineno{#1*}}%
\AtBeginDocument{%
  \patchBothAmsMathEnvironmentsForLineno{equation}%
  \patchBothAmsMathEnvironmentsForLineno{align}%
  \patchBothAmsMathEnvironmentsForLineno{gather}%
  \patchBothAmsMathEnvironmentsForLineno{multline}%
  \patchBothAmsMathEnvironmentsForLineno{flalign}%
  \patchBothAmsMathEnvironmentsForLineno{alignat}%
}
\newtcolorbox{mynote}[1][]{
  colback=blue!5!white,
  colframe=gray!75!black,
  fonttitle=\bfseries,
  title=Note,
  arc=1mm,
  #1
}
\hypersetup{colorlinks=true,linkcolor=blue,citecolor=blue,urlcolor=blue}
\lstdefinestyle{pythonstyle}{
    language=Python,
    backgroundcolor=\color[rgb]{0.95,0.95,0.95},
    commentstyle=\color[rgb]{0,0.5,0},
    keywordstyle=\color{blue},
    stringstyle=\color[rgb]{0.8,0.4,0},
    basicstyle=\footnotesize\ttfamily,
    breakatwhitespace=false,
    breaklines=true,
    captionpos=b,
    keepspaces=true,
    numbers=none,
    numbersep=5pt,
    showspaces=false,
    showstringspaces=false,
    showtabs=false,
    tabsize=2,
    frame=single,
    rulecolor=\color{black!30},
}

\title{netseg: a Python Package for Measuring Structural Polarization and Segregation in Social Networks}
\author{Onur Tuncay Bal\textsuperscript{1} \and Micha\l{} Bojanowski\textsuperscript{2}}
\date{\today}
\begin{document}
\makeatletter
\twocolumn[
  \begin{@twocolumnfalse}
    \maketitle
    \begin{center}
      \begin{minipage}{0.85\textwidth}
        \begin{center}\textbf{Abstract}\end{center}
        \small\noindent The study of structural polarization and segregation in social networks is an established line of research, and the quantification of both phenomena proceeds through a set of widely cited network indices. The code implementing those indices, however, is seldom released and almost never tested. We present netseg, a comprehensively documented Python package implementing these indices, most of them generalized to more than two groups and to directed as well as undirected input. It ports the R package of the same name and adds measures the R version lacks, among them Random Walk Controversy, Boundary Connectivity, Dipole Moment, and Moran's I. The package operates on igraph objects and performs the underlying graph operations (e.g., neighborhood queries and random-walk simulation) through igraph's Python interface, so that they execute in compiled code rather than in interpreted Python. For several of the indices this yields runtimes orders of magnitude below those of the available open-source implementations, which the documentation reports in benchmarks. Most of these indices are defined as a divergence from a null model, and published implementations fix that null model to a uniform random graph of matching density. netseg accepts an ensemble of graphs as a sample from an arbitrary null model, and distinguishes indices that already incorporate a baseline from those that do not, adjusting the comparison accordingly to avoid double subtraction. The documentation provides, for each index, a worked empirical example, its behaviour at the degenerate cases where it is undefined, a benchmark, and the procedure for substituting a custom null model. We report the behaviour of every index over a parameter sweep of a generative opinion model, and apply them to a county-level railroad network built from nineteenth-century operator records joined to full-count census data.
      \end{minipage}
    \end{center}
    \vspace{1.5\baselineskip}
  \end{@twocolumnfalse}
]
\makeatother

\footnotetext[1]{Department of Network and Data Science, Central European University, Vienna, Austria}
\footnotetext[2]{Department of Quantitative Methods and Information Technology, Kozminski University, Warsaw, Poland and Department of Social and Cultural Anthropology, Autonomous University of Barcelona, Spain}
\section*{Summary}

Polarization denotes the structural bifurcation of a population into diametrically opposed subgroups, communities, or ideological paradigms \citep{converse1964nature, interian2023network, stokes_1999}. Analogously, segregation defines the degree of relational or spatial separation maintained between distinct groups \citep{bojanowski2014measuring, freeman1978segregation, massey1988dimensions}. Though sometimes conflated in the literature, the manifestation of both phenomena within social networks facilitate network fragmentation along ideological axes, precipitating a range of detrimental socio-political outcomes \citep{barbera2020social}. Previous studies have found that network polarization and segregation hamper the diffusion of information \citep{halberstam2016homophily}, reinforce confirmation bias through the formation of echo chambers \citep{ sunstein2018republic}, or undermine democratic discourse \citep{barbera2020social, beaufort2018digital, bright2018explaining, keijzer2026computational}.

Given the severity of these outcomes, an interdisciplinary effort has emerged to better define, quantify, and mitigate both phenomena. Because polarization and segregation are inherently relational, network science provides a uniquely suited framework for their analysis. The rapid adoption of this approach is evidenced by its bibliometric footprint, a sweep of the OpenAlex database, with topics related to network science and computational social science, identifies roughly 8,000 English-language works indexing ``network'' and ``polarization'' in their title or abstract published between 2010 and 2026.\footnote{The sweep can be reproduced with the ``.ipynb'' \href{https://codeberg.org/OnurB/netseg/raw/branch/main/open_alex_audit.ipynb}{file} within the \href{https://codeberg.org/OnurB/netseg}{repository}.}

However, while several foundational metrics remain widely used, recent scholarship has begun to interrogate their validity and propose necessary extensions. This includes using varying null models \citep{salloum2022separating} to compare existing measures and evaluate how specific tie-formation micro-mechanisms explain overall segregation and structural polarization \citep{duxbury_micro_2023}.

\section*{Statement of need}

\texttt{netseg}\footnote{The package is hosted in \href{https://codeberg.org/OnurB/netseg}{here.}} is a Python package for measuring structural polarization and segregation in social networks. It is not only an adaptation of the existing R package \texttt{netseg} \citep{bojanowski2026package} to the Python environment, but a substantial extension involving state of the art structural polarization measures and custom null-model support with extensive \href{https://onurb.codeberg.page/netseg/}{documentation}. Although the quantification of segregation and structural polarization phenomena has become standard practice, the lack of a robust, easy-to-implement computational framework leaves a persistent methodological gap. While foundational algorithms are routinely cited, their practical implementation has become a severe operational bottleneck for researchers.

As shown in previous studies \citep{soergel2015rampant, ziemann2016gene}, the absence of validated software can have a deleterious effect. Since the code governing these metrics is rarely published in open repositories and undergoes no formal testing, comparison and reproducibility become challenging. So much so that out of the 8,000 works mentioned previously, less than 2\% contain a repository linked to the paper's metadata. To solve this issue, \texttt{netseg} offers a unified toolkit that is extensively tested for both the Python and R programming languages, making cross-comparison between studies possible even across different platforms.

Beyond being ready ``out of the box'', \texttt{netseg} is also implemented with scalability in mind. Some of the previous individual implementations of the metrics (even when they are implemented in parallel) suffer from Python's interpreter overhead. \texttt{netseg} avoids this completely using vectorized operations that are built on already optimized libraries such as \texttt{scipy} and \texttt{NumPy}. Furthermore, some of the previous implementations use \texttt{networkx}, a library that, by its design principle, employs Python dictionaries. \texttt{netseg} relies on \texttt{igraph}'s C backend \citep{igraph2023, igraph2006} for operations such as neighborhood detection or Monte Carlo simulations. As a result, the quantification of segregation and structural polarization on social networks formed around big data becomes not just faster to analyse, but possible. Written in the two most popular languages for data and network analysis, \texttt{netseg} offers extensive documentation. The documentation demonstrates each metric in detail with empirical examples, while the package itself offers detailed input validation and docstrings.

Finally, most segregation and structural polarization metrics rely on a random graph with the same edge density for comparative analysis, a notoriously low bar to pass. \texttt{netseg} allows supplying custom null models as graph ensembles directly within the function calls, resulting in a more robust comparison. Additionally, \texttt{netseg} extends most of these metrics to accommodate multiple groups and multimode networks, detailing these extensions explicitly within the documentation. This allows users to execute complex, multi-group analyses without requiring bespoke custom scripting.

\section*{State of the field}

The most direct predecessor of the present package is the R package \texttt{netseg} \citep{bojanowski2026package}, which implements the classical segregation and homophily indices and accepts network data either as \texttt{igraph} objects or, for those measures where it is sufficient, as a mixing matrix. Its canonical reference remains the substantive article it accompanies \citep{bojanowski2014measuring}, which has itself accumulated over one hundred citations. Since its first CRAN release in February 2021 it has been downloaded more than 18,000 times from the Posit CRAN mirror alone. The rate has not decayed, with roughly 3,700 downloads in the twelve months to August 2026.\footnote{CRAN download counts obtained from the \texttt{cranlogs} service; repository statistics from the source repository at \url{https://github.com/mbojan/netseg}. All figures accessed 1 September 2026.} The package remains maintained, at version 1.0-3 as of April 2025, carries 20 stars on GitHub, and is listed in the CRAN \emph{NetworkAnalysis} Task View among the tools for exploratory analysis of networks. Task Views are guides into R packages in specific fields of research or applied work curated by experts.

The character of that usage matters as much as its volume. Publicly available code calling \texttt{netseg} appears in replication packages for published research \citep{estevez2024denunciation}, in doctoral and capstone projects, in studies of adolescent friendship and of situated mental health, and in graduate social network analysis course materials; the R package \texttt{netmediate} \citep{duxbury_micro_2023} documents \texttt{netseg::assort()} as its worked example of a user-supplied macro-level statistic. Research using \texttt{netseg} appeared in prestigous journals such as \textit{Science Magazine} \citep{lee_etal_2023} or \textit{Nature} \citep{simpson_etal_2023}. Researchers working in Python, however, cannot reach any of it without friction. Bridging to R from Python is possible through \texttt{rpy2}, but it requires provisioning and maintaining an R installation alongside the Python environment, converting graph objects across the language boundary on every call, and it forfeits the vectorized execution path described below. In practice the cost of that bridge pushes Python users toward reimplementing individual metrics by hand, with exactly the reproducibility consequences discussed in the previous section.

Beyond resolving these language barriers, the Python package expands upon the original toolset. Although the R package contains the majority of the classical metrics, it still lacks a few contemporary measures such as Random Walk Controversy (RWC), Boundary Connectivity, Dipole Moment, as well as Moran's I, which is standard in spatial analysis but rarely applied to attribute segregation on networks. Furthermore, the Python package introduces custom null model support. Users can change each metric's null model internally, or compare a metric to a null model by supplying an ensemble of graphs as a list to the function calls. This additional development establishes the Python package not just as a port, but as an extended and contemporary application of the R package.

While a few prior implementations of these metrics exist, they are generally not comprehensive implementations. Instead, they tend to be standalone scripts or isolated functions within monolithic libraries. For instance, a parallel implementation of the Random Walk Controversy (RWC) measure is available on PyPI as \texttt{random-walk-controversy}, which relies directly on \texttt{networkx} for its Monte Carlo simulations of random walks required for the metric. For common metrics such as \emph{assortativity}, generalized libraries like \texttt{networkx} and \texttt{igraph} provide built-in functions; however, for specialized measurements like boundary connectivity and dipole moment, researchers must often resort to repositories without documentation \citep{salloum2022separating} or that contains outdated scripts that belongs to the original authors of the metrics \citep{garimella2018quantifying}. Another tool that can be used for quantifying segregation is the established social network analysis software UCINET \citep{guide1999ucinet}, but it is GUI-based and, according to its documentation, it is limited to a maximum of 32,767 nodes. Finally, NetIn \citep{pynetin} is a package written over \texttt{networkx} for quantification and analysis of network inequalities, differing from quantification of segregation and structural polarization, while forming around similar concepts.

\section*{Research impact statement}

The introduction of \texttt{netseg} significantly impacts the scalability and robustness of analysing segregation and structural polarization in social networks. By offering a unified, extensively tested toolkit available in both Python and R, the package establishes a standard for measuring structural polarization and segregation. It leverages vectorized operations and the C-backed architecture of \texttt{igraph} to eliminate interpreter overhead, enabling the analysis of large-scale social networks that were previously computationally prohibitive.

One major methodological contribution of \texttt{netseg} is its transformative impact on null-model comparison approaches. Most of these metrics are defined as divergences from a null model, and in nearly all published implementations that null model is a uniform random graph of matching density. \texttt{netseg} advances the approach by allowing researchers to supply custom graph ensembles — interpreted as samples from a null model, directly within function calls. In practice, these ensembles might be networks with preserved degree distributions or counterfactual simulations obtained from other generative statistical models (e.g., Exponential Random Graph Models). The package distinguishes between metrics that inherently contain a null model (e.g., Boundary Connectivity) and those that do not (e.g., Random Walk Controversy), automatically adjusting baselines or calculating excess segregation while strictly preventing double subtraction errors.

To maximize its utility and accessibility, \texttt{netseg} is thoroughly documented with comprehensive, empirical examples. The documentation explicitly details how to implement these custom null models through code snippets, demonstrating how users can evaluate counterfactual ``what-if'' scenarios to isolate the impact of specific network mechanisms on final segregation scores. Furthermore, the package transparently addresses the computational nuances of these advanced analyses within its examples.

Currently \texttt{netseg} extends and incorporates the metrics summarized in Table~\ref{tab:metrics}.

\section*{Installation and Documentation}

\texttt{netseg} can be installed directly from PyPI with 

\begin{lstlisting}[style=pythonstyle]
pip install netseg
\end{lstlisting}
or the latest version can be installed from the source with 
\begin{lstlisting}[style=pythonstyle]
git clone --filter=blob:none --sparse https://codeberg.org/OnurB/netseg.git
cd netseg
git sparse-checkout set src
pip install .
\end{lstlisting}

The package documentation contains all of the functions, their use cases, their edge case behaviours and analysis of the metrics in detail. Furthermore, it also contains benchmarks, general behaviour of the metrics and their application on empirical datasets. 

The documentation is hosted on \href{https://onurb.codeberg.page/netseg/}{Codeberg Pages}. We use this service for ethical reasons. Nonetheless, it may occasionally experience downtime. If the site is unavailable, you can always read the compiled HTML documentation locally via the \texttt{pages} branch:

\begin{lstlisting}[style=pythonstyle]
git clone --branch pages --single-branch --depth 1 https://codeberg.org/OnurB/netseg.git netseg-docs
cd netseg-docs
python3 -m http.server 8000
\end{lstlisting}
Then go to localhost:8000 on your browser.

\section*{Design and Usage}

Designed with simplicity and scalability in mind, \texttt{netseg} can be used in various ways. First usage case, and maybe the most common especially within sociology is to supply a mixing matrix directly. This method can be used with \texttt{netseg} if the underlying quantification method relies on the mixing matrix solely.

\begin{lstlisting}[style=pythonstyle]
from netseg import freeman
import numpy as np

membership = [0 if i < 100 else 1 for i in range(200)]
mm = np.array([[140, 16],[16,140]])
freeman(membership= membership,
        mixing_matrix = mm)
>> 0.79
\end{lstlisting}

This approach expects a mixing matrix as a \texttt{np.array} and also a membership \texttt{list}. Notice that membership \texttt{list} is necessary to calculate the expected number of edges forming within the group.

The package can also be used with an \texttt{igraph.Graph} object. In this case, \texttt{netseg} constructs the mixing matrix itself efficiently with \texttt{scipy} sparse matrix multiplication, avoiding the interpreter overhead completely. We can demonstrate it on a random graph as follows.

\begin{lstlisting}[style=pythonstyle]
import igraph as ig
g = ig.Graph.Erdos_Renyi(200, 0.01)
membership = [0 if i < 100 else 1 for i in range(g.vcount())]
freeman(membership = membership,
        graph = g)
>> np.float64(0.0722)
\end{lstlisting}

Finally, \texttt{netseg} can be used with vertex attributes of an \texttt{igraph.Graph} object directly if the attribute \texttt{str} is supplied as a parameter to any function that expects a \texttt{membership} argument.

\begin{lstlisting}[style=pythonstyle]
g.vs['group_membership'] = membership
freeman(membership = 'group_membership',
        graph = g)
>> np.float64(0.0722)
\end{lstlisting}

\section*{Example on a Generative Statistical Model}

To illustrate how the metrics behave and how \texttt{netseg} operates, we rely on a simple generative model driven by the opinions of the agents. We begin by defining an opinion space as a vector in which each element encodes the opinion of one agent on a single issue within a population. Individual opinions are drawn from a Gaussian mixture model, which lets us regulate the level of polarization by adjusting the means and variances of the component distributions. Formally, the opinion space is a vector $X \in \mathbb{R}^n$, where each of the $n$ agents holds a stance value between $-1$ and $1$ on the topic. To reproduce opinion diversity and polarization, the values of $X$ are drawn independently from a two-component Gaussian mixture with means $\mu_1$, $\mu_2$ and standard deviations $\sigma_1$, $\sigma_2$:

\begin{equation*}
X_i \sim \pi\,\mathcal{N}(\mu_1, \sigma_1^2) + (1-\pi)\,\mathcal{N}(\mu_2, \sigma_2^2),
\quad i = 1,\dots,n.
\end{equation*}
Draws falling outside $[-1,1]$ are clipped to the interval boundary, so that $X \in [-1,1]^n$. 

This grants direct control over both the separation of opinions (through $\mu_1, \mu_2$) and their spread (through $\sigma_1$, $\sigma_2$). The model accommodates varying levels and asymmetries of polarization, but we fix $|\mu_1| = |\mu_2|$ for simplicity. Taken together, these parameters govern the overall shape of the distribution and let us simulate different levels of ideological polarization. We set $\pi = 0.5$ throughout, so the two groups are of equal expected size.

With the opinion space in place, we can start building the network itself. We parameterize our model with the following:

\begin{equation*}
\log \left( \frac{P_{ij}}{1 - P_{ij}} \right) = \alpha + \beta_1 \cdot |x_i - x_j| + \beta_2 \cdot (x_i + x_j)
\end{equation*}
\begin{equation*}
P_{ij} = \frac{1}{1 + \exp\left(-\left(\alpha + \beta_1 \cdot |x_i - x_j| + \beta_2 \cdot (x_i + x_j)\right)\right)}
\end{equation*}

Where $x_k$ denotes the opinion score of node $k$. The parameter $\alpha$ establishes the baseline network density. The coefficient $\beta_1$ adjusts homophily such that lower values of $\beta_1$ penalize the absolute distance $|x_i - x_j|$, reducing the probability of ties between dissimilar nodes. Finally, $\beta_2$ adjusts degree asymmetry across the opinion spectrum; a positive $\beta_2$ increases the expected degree of nodes with positive opinion scores, and vice versa. Note that each node's group membership is determined by its originating mixture component: a node whose opinion is drawn from component 1 is assigned to group 1, and likewise for component 2.

We can demonstrate the behaviour of the metrics implemented in \texttt{netseg}. We generate 1,600 networks with varying parameter combinations. We setup the parameter space as 40 to 40 grid where $\beta_1$ takes values in between $-4$ and $4$ while $\beta_2$ takes values in between $0$ and $0.6$ (negative values will change the group with total higher degree, so for demonstration it is redundant).

Figure~1 illustrates the values metrics take within this parameter space. Notice that the if the cell contains an empty value, generated graph is not satisfying the logical prerequisites for the metric (e.g., lack of boundary nodes for boundary connectivity in case of graphs with no cross-group ties).

\section*{Example Usage on Empirical Data}

To demonstrate the package on empirical data, we construct a county-level railroad network obtained from historical data \citep{atack2015railroads} and attach county-level African American population ratios drawn from the IPUMS full-count census \citep{ruggles2024fullcount}. The railroad shapefile stores each track segment together with the company operating it, recorded in the \texttt{InOpBy} attribute. We treat counties as vertices and connect two counties whenever a single operator runs track through both, so that shared operational control, rather than geographic adjacency, generates the ties.

The construction proceeds in three steps. First, the line geometry is simplified with the Douglas-Peucker algorithm at a 2 meter tolerance, which removes redundant vertices while preserving topology. Second, the segments are grouped by operator and merged on a 1 km grid, and a spatial index records which counties each merged component intersects. Third, for every operator, all pairs of intersected counties receive an edge labelled with that operator, and parallel edges are removed with \texttt{simplify()}. Figure~\ref{fig:trainsimplified} contrasts the raw operator geometry with the resulting abstract network.

Each vertex carries a binary attribute \texttt{BRATIO} that marks whether the county's African American population ratio falls at or above the sample median. We compute the Spectral Segregation Index for every vertex directly from the attribute name:

\begin{lstlisting}[style=pythonstyle]
from netseg import ssi

railroad_graph.vs['ssi_score'] = ssi(
    'BRATIO',
    railroad_graph,
    aggregate=False)
\end{lstlisting}

Passing \texttt{aggregate=False} returns one score per vertex instead of a single group-level summary. We then map these scores onto the network layout in Figure~\ref{fig:ssi_output}. Vertices embedded in neighborhoods dominated by their own group take higher values, which locates the most segregated pockets of the operator-linked county structure.

\section*{AI usage disclosure}

The authors utilized Generative AI to assist with specific aspects of the package's documentation and testing for Python code. Specifically, AI was used to generate boilerplate NumPy-style docstrings based on provided functions, parameters, and their intended descriptions. Following the creation of test templates, the AI also assisted in writing input validation tests (All of the metric measurement tests are done manually, if applicable while ensuring \texttt{netseg} returns the exact result given in the original paper). Within the core Python code development (package itself), the use of Generative AI was strictly limited to debugging and identifying potential edge cases to ensure robust error handling.

\bibliography{references}

\onecolumn

\section*{Tables}

\begin{table}[h!]
\centering
\footnotesize
\caption{Metrics implemented in \texttt{netseg}: network type support, multi-group extensibility, computational complexity, and value range. $M$ = number of null-models (if none supplied $M = 1$), $V$ = nodes, $E$ = edges, $K$ = groups, $S$ = number of walks, $L$ = walk length, $I$ = iterations. \texttt{netseg} relaxes the edge-orientation constraints of classical segregation indices except Spectral Segregation Index and also generalizes the metrics for multiple groups if applicable. RWC complexity excludes one-off cost of seed selection.}
\label{tab:metrics}
\begin{tabularx}{\textwidth}{@{}L L l l l L L@{}}
\toprule
Metric & Author & Orig. type & \texttt{netseg} type & Multi-group & Complexity & Range \\
\midrule
Boundary Connectivity & \citet{guerra2013measure} & Undir. & Both & Yes & $O(M(V+E))$ & $[-0.5, 0.5]$ \\
Random Walk Controversy & \citet{garimella2018quantifying} & Undir. & Both & Yes & $O(M \cdot S \cdot L)$ & $[-1, 1]$ \\
Dipole Moment & \citet{morales2015measuring} & Undir. & Both & Two only & $O(M \cdot I \cdot E)$ & $[0, 1]$ \\
Freeman's Seg. Index & \citet{freeman1978segregation} & Undir. & Both & Yes & $O(M(V+E))$ & $[0, 1]$ \\
Krackhardt E-I Index & \citet{krackhardt1988informal} & Both & Both & Yes & $O(M(V+E))$ & $[-1, 1]$ \\
Segregation Matrix Index & \citet{fershtman1997cohesive} & Dir. & Both & Yes & $O(M(V+E))$ & $[-1, 1]$ \\
Coleman's Homophily Index & \citet{coleman1958relational} & Dir. & Both & Yes & $O(M \cdot K(V+E))$ & $[-1, 1]$ \\
GAM Index & \citet{gupta1989networks} & Both & Both & Yes & $O(M(V+E+K^2))$ & $\left[-\frac{1}{K-1}, 1\right]$ \\
ORWG & \citet{moody2001race} & Both & Both & Yes & $O(M(V+E))$ & $[0, \infty)$ \\
Spectral Segregation Index & \citet{echenique2007measure} & Undir. & Undir. & Yes & $O(V + M \cdot V^3)$ & $[0, \infty)$ \\
Assortativity Coefficient & \citet{newman2003mixing} & Both & Both & Yes & $O(M(V+E))$ & $\left[-\frac{\sum a_i b_i}{1 - \sum a_i b_i}, 1\right]$ \\
Moran's I & \citet{moran1950notes} & Both & Both  & Yes & $O(M(V+E))$ & $[-1, 1]$ \\
\bottomrule
\end{tabularx}
\end{table}

\clearpage

\section*{Figures}

\begin{figure}[H]
    \centering
    \includegraphics[width=0.85\linewidth]{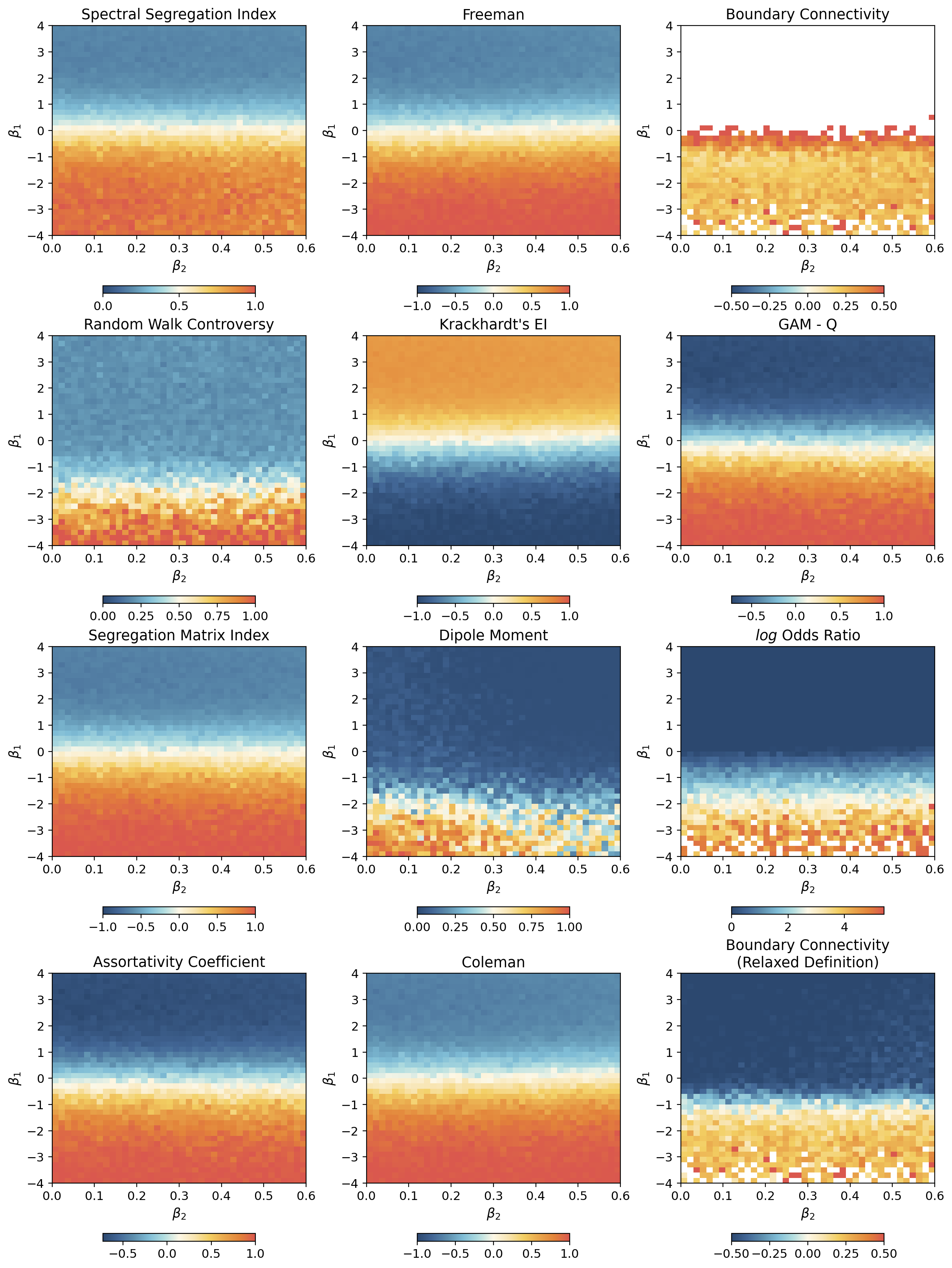}
    \caption{Values of each metric implemented in \texttt{netseg} across the generative model parameter space. Every panel reports one metric evaluated over the $40 \times 40$ grid of $(\beta_1, \beta_2)$ combinations, with $\beta_1 \in [-4, 4]$ and $\beta_2 \in [0, 0.6]$. Empty cells mark parameter combinations whose generated graph does not satisfy the logical prerequisites of the corresponding metric.}
    \label{fig:behaviour}
\end{figure}

\clearpage

\begin{figure}[H]
    \centering
    \includegraphics[width=0.99\linewidth]{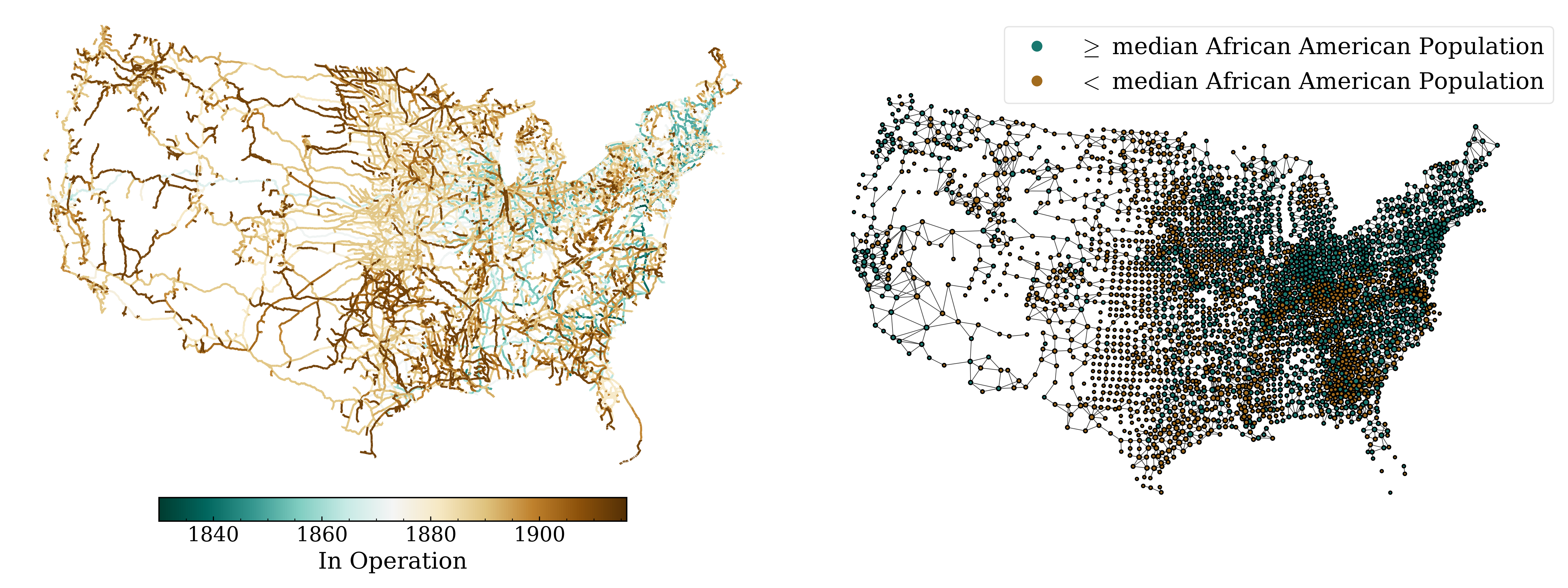}
    \caption{Left: raw railroad segments from Jeremy Atack's data, colored by operating company (\texttt{InOpBy}). Right: the abstracted county network, in which two counties are joined when a common operator runs track through both. Vertex size scales with degree, and vertex color separates counties at or above the median African American population ratio from those below it.}
    \label{fig:trainsimplified}
\end{figure}

\begin{figure}[H]
    \centering
    \includegraphics[width=0.99\linewidth]{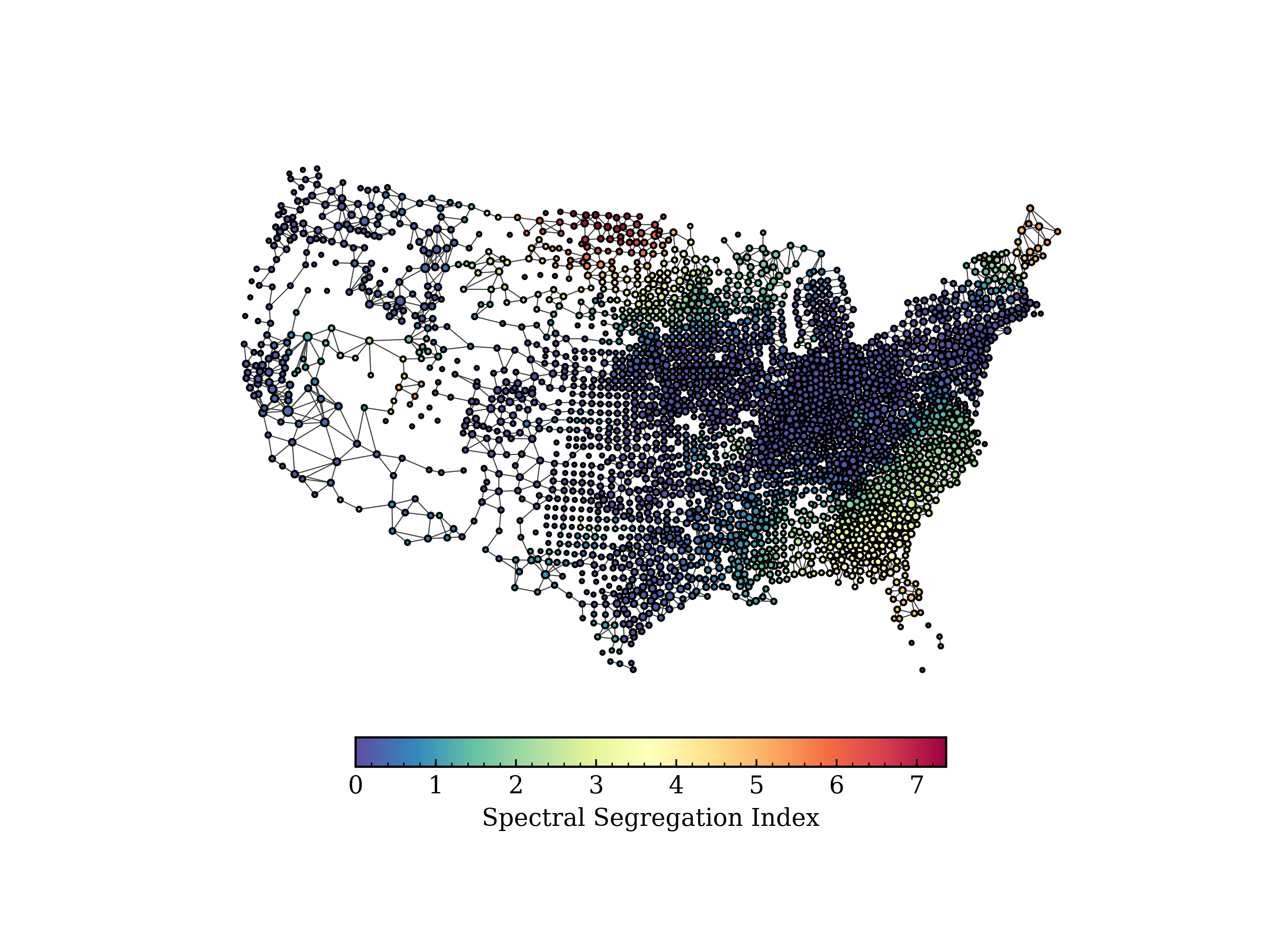}
    \caption{The county railroad network with vertices colored by their per-vertex Spectral Segregation Index, computed on the binary median African American population ratio attribute. Higher values identify vertices embedded in more segregated local neighborhoods.}
    \label{fig:ssi_output}
\end{figure}

\end{document}